\documentclass{elsart}

\usepackage{natbib}

 \usepackage{graphicx}
\usepackage{amssymb,amsmath,alltt}
\newcommand{\grad}{\nabla}

\newcommand{\bk}{\boldsymbol{k}}
\newcommand{\bx}{\boldsymbol{x}}
\newcommand{\bdx}{\boldsymbol{dx}}
\newcommand{\bforce}{\boldsymbol{f}}
\newcommand{\bvel}{\boldsymbol{v}}

\begin{document}

\begin{frontmatter}

\title{Out-of-core Hydrodynamic Simulations for Cosmological Applications}




\author[Astro,CITA]{Hy Trac},
\ead{trac@cita.utoronto.ca}
\author[CITA]{Ue-Li Pen}
\ead{pen@cita.utoronto.ca}
\address[Astro]{Department of Astronomy and Astrophysics, University of Toronto, Toronto, ON M5S 3H8, Canada}
\address[CITA]{Canadian Institute for Theoretical Astrophysics, 60 St. George Street, Toronto, ON M5S 3H8, Canada}

\begin{abstract}
We present an out-of-core hydrodynamic code for high resolution cosmological simulations that require terabytes of memory.  Out-of-core computation refers to the technique of using disk space as virtual memory and transferring data in and out of main memory at high I/O bandwidth.  The code is based on a two-level mesh scheme where short-range physics is solved on a high-resolution, localized mesh while long-range physics is captured on a lower resolution, global mesh.  The two-level mesh gravity solver allows FFTs to operate on data stored entirely in memory, which is much faster than the alternative of computing the transforms out-of-core through non-sequential disk accesses.  We also describe an out-of-core initial conditions generator that is used to prepare large data sets for cosmological simulations.  The out-of-core code is accurate, cost-effective, and memory-efficient and the current version is implemented to run in parallel on shared-memory machines.  I/O overhead is significantly reduced down to less than 10\% by performing disk operations concurrently with numerical calculations.  The current computational setup, which includes a 32 processor Alpha server and a 3 TB striped SCSI disk array, allows us to run cosmological simulations with up to $4000^3$ grid cells and $2000^3$ dark matter particles.
\end{abstract}

\end{frontmatter}

% main text

\section{Introduction}

Presently, one of the big tasks in cosmology is to determine the concordance model motivated by the results from the cosmic microwave background (CMB), large-scale structure (LSS), and supernovae.  The precision measurement of the matter power spectrum is one key scientific goal and observations of the Lyman alpha (Ly$\alpha$) forest, the Sunyaev-Zeldovich (SZ) effect, and weak lensing of the LSS are expected to complement existing constraints.  In order to do cosmology with these probes, numerical modelling of the nonlinear physics must achieve a level of accuracy on par with upcoming precision observations.  

Cosmological hydrodynamic and N-body simulations are standard tools for modeling nonlinear structure formation in the universe.  Numerical simulations must converge over a large range in scale, mass, and temperature.  Large box sizes are required to capture large-scale correlations and power while high resolution is needed to resolve small-scale, nonlinear structures.  Most simulations to date have sacrificed one for the other because of the limitations in computational resources.  While parallelized numerical codes and optimized mathematical libraries have significantly reduced computation time on high performance computing (HPC) systems, memory limitations remain the bottleneck in expanding the size of simulations.

We describe the implementation of an out-of-core hydrodynamic (OCH) cosmological code for high resolution simulations requiring terabytes of memory.  Out-of-core computation refers to the technique of using disk space as virtual memory and transferring data in and out of main memory at high I/O bandwidth.  Conventional memory is an expensive commodity and currently most HPC systems have a few hundred gigabytes at most.  However, disk space is relatively cheap and disk arrays can provide many more orders of magnitude in storage.  In order to be effective, out-of-core computation must avoid being I/O limited.  Striped SCSI disk arrays are capable of delivering on the order of $100-1000$ MB/s throughput and have sufficient bandwidth.  However, the slow latency presents a major problem for codes requiring non-sequential disk access.  Pioneering work has been done for computational astrophysics.  \citet{1997SalmonWarren} developed a parallel, out-of-core tree N-body code to run an 80 million particle simulation on a distributed-memory Pentium cluster with a total of 16 processors, 2 GB of memory, and 16 GB of disk space.  The out-of-core paradigm remains a niche that has been relatively unexplored for numerical simulations.  We demonstrate that advances in algorithmic development now make it feasible to do out-of-core computation effectively.

The OCH code is designed for cosmological applications where high mass resolution is required at all scales.  The code is based on the Hydro\&N-body implementation of \citet{2004TracPenMACH}.  The hydrodynamics of the baryonic gas is captured using an Eulerian total variation diminishing (TVD) scheme \citep{1983Harten} that provides high resolution capturing of shocks while preventing unphysical instabilities.  The gravitational evolution of the dark matter is tracked using the standard particle-mesh (PM) scheme \citep{1988HockneyEastwood}.  

The challenges in doing out-of-core computation stem from the requirement that the data domain be decomposed into blocks which can fit in memory.  The hydrodynamics of the ideal gas is a local process that is straightforward to solve on the decomposed domain with the addition of buffer regions.  However, gravity is a global force and Poisson solvers operate on the global density field.  Fast Fourier transforms (FFTs) are the optimal solvers in a standard PM scheme, but they involve global transposes and computing the transforms out-of-core will be intolerably slow.  This problem can be addressed by splitting the gravitational force into long and short-range components, similar to that done in P$^3$M \citep{1988HockneyEastwood}, mesh-refined PM \citep{1991Couchman, 1995CouchmanHydra}, and Tree-PM \citep{1995Xu, 2002Bagla, 2003BodeOstrikerTPM, 2004DubinskiGOTPM} codes.  We implement a two-level mesh scheme where the short-range force is solved on a high-resolution, localized mesh while the long-range force is captured on a lower resolution, global mesh.  The two-level mesh gravity solver is memory-efficient and allows FFTs to be performed on data stored entirely in memory.  

In this paper, we describe the out-of-core algorithm and highlight the steps required for doing out-of-core computation.  In particular, we discuss optimizations to reduce I/O overhead by performing disk operations and numerical calculations concurrently.  The two-level mesh gravity solver is described and its accuracy is compared to that of a standard one-level mesh solver.  In addition, we present an out-of-core initial conditions generator that can be used to construct large data sets for high resolution cosmological simulations.

\section{Out-of-Core Algorithm}
\label{sec:algorithm}

The three-dimensional out-of-core hydro (OCH) code is based on the Hydro\&N-body implementation of \citet{2004TracPenMACH}.  For reference, a pedagogical primer on Eulerian computational fluid dynamics is presented in \citep{2003TracPenCFD}.  We briefly describe the cosmological code and present modifications specifically added for doing out-of-core computation.  The code is designed for cosmological applications in an Friedman-Robertson-Walker (FRW) universe.

The out-of-core implementation requires both the division of data into blocks that can fit in memory and the decomposition of physical laws with global dependence into short and long range components.  The code uses a two-level mesh algorithm where short-range physics is solved on a high-resolution, localized mesh while long-range physics is captured on a lower resolution  global mesh.  The global mesh is 4 times coarser in each dimension, reducing memory usage by a factor of 64 when calculating the long-range interactions.  The standard technique of having buffer regions around the local blocks ensures that the short-range interactions are calculated with no boundary effects imposed by the domain decomposition.  In the sections to follow, we describe the Hydro\&N-body implementation and the two-level mesh gravity solver.

\subsection{Hydro\&N-body}

The hydrodynamics of the baryonic gas is solved using an Eulerian algorithm based on the moving frame approach of \citet{2004TracPenMACH} and the second-order accurate total variation diminishing (TVD) scheme \citep{1983Harten}.  The Eulerian conservation equations are solved in an adaptive frame moving with the fluid where local fluid variables can be directly tracked.  The moving frame approach minimizes the numerical Mach numbers, allowing high resolution capturing of shocks and preventing spurious oscillations in both subsonic and supersonic regions.  Time integration is performed using a second-order accurate Runge-Kutta technique.  The moving frame hydro algorithm uses an operator-splitting technique \citep{1968Strang} to efficiently solve the various flux terms in the Euler equations.  To get second-order accuracy, each hydro iteration is composed of a double time step and the order of operations is reversed in the second time step.  A single gravity step is done between the forward and reverse hydro steps.

The gravitational evolution of the dark matter particles is tracked using the standard PM N-body scheme \citep{1988HockneyEastwood}.  We build upon the implementation of \citet{2004TracPenMACH}, which is based on a single-mesh gravity solver.  In the two-level mesh scheme, mass assignment and force interpolation on both the fine and coarse grids are accomplished using `cloud-in-cell' (CIC) interpolation.  Cubical cloud shapes resembling the fine or coarse cells are used for the respective grids.  The collisionless particles are advected by solving Newton's equations of motion in an FRW universe.  Time integration is performed using an explicit, second-order accurate Runge-Kutta technique that allows synchronization between the dark matter particles and the gas.

\subsection{Gravity}

\subsubsection{Two-level mesh gravity solver}

Gravity is a global force that can be decomposed into short and long range components and solve with a two-level mesh scheme.  Poisson's equation
\begin{equation}
\grad^2\phi=4\pi G\rho,
\end{equation}
relates the gravitational potential $\phi$ to the density field $\rho$ and the general solution can be written as
\begin{equation}
\phi(\bx)=\int\rho(\bx')w(\bx-\bx')d^3x',
\end{equation}
where the isotropic kernel is given by
\begin{equation}
\label{eqn:globalpotential}
w(r)=-\frac{G}{r}.
\end{equation}
In the standard mesh approach, the convolution is done in Fourier space and fast Fourier transforms (FFTs) are used to convert the discrete density field and recover the discrete potential field.  Since FFTs are highly nonlocal and involve global transposes, computing the transforms out-of-core using non-sequential disk accesses will be intolerably slow.  However, the two-level mesh scheme allows us to apply the FFTs to data that are entirely stored in memory.

We decompose the potential kernel $w(r)$ into a short-range component
\begin{equation}
w_s(r)=
\begin{cases}
w(r)-\alpha(r) & \text{if $r\leq r_c$},\\
0 & \text{otherwise},
\end{cases}
\label{eqn:ws}
\end{equation}
and a long-range term
\begin{equation}
w_l(r)=
\begin{cases}
\alpha(r)\ \ \ \ \ \ \ \ \ \ & \text{if $r\leq r_c$},\\
w(r) & \text{otherwise},
\end{cases}
\label{eqn:wl}
\end{equation}
where the short-range cutoff $r_c$ is a free parameter.  The function $\alpha(r)$ is chosen to be a polynomial,
\begin{equation}
\alpha(r)=G(a+br^2+cr^4),
\end{equation}
whose coefficients, 
\begin{equation}
\begin{aligned}
a&=-\frac{15}{8r_c},\\
b&=\frac{10}{8r_c^3},\\
c&=-\frac{3}{8r_c^5},
\end{aligned}
\end{equation}
are determined from the conditions
\begin{equation}
\label{eqn:kernelconditions}
\begin{aligned}
\alpha(r_c)&=w(r_c),\\
\alpha^\prime(r_c)&=w^\prime(r_c),\\
\alpha^{\prime\prime}(r_c)&=w^{\prime\prime}(r_c).
\end{aligned}
\end{equation}
These restrictions ensure that the long-range kernel smoothly turns over near the cutoff and that the short-range term smoothly goes to zero at the cutoff.

The long-range potential $\phi_l^c(\bx)$ is computed by performing the convolution over the coarse-grained global density field $\rho^c(\bx)$.  The superscript $c$ denotes that the discrete fields are constructed on a coarse grid.  The total density field has contributions from both the dark matter particles and baryonic gas cells.  Mass assignment onto the coarse grid is accomplished using the `cloud-in-cell' (CIC) interpolation scheme \citep{1988HockneyEastwood} with cloud shape being the same as a coarse cell.  The long-range force field $\bforce_l^c(\bx)$ is obtained by finite differencing the long-range potential and force interpolation  is carried out using the same CIC scheme to ensure no fictitious self-force.

Since the two-level mesh scheme uses grids at different resolutions, the decomposition given by equations (\ref{eqn:ws}) and (\ref{eqn:wl}) needs to be modified.  In Fourier space, we can write the long-range potential as
\begin{equation}
\tilde{\phi}_l(\bk)=\tilde{\rho}^c(\bk)\tilde{w}_l^c(\bk)=[\tilde{\rho}(\bk)\tilde{s}_\rho(\bk)][\tilde{w}_l(\bk)\tilde{s}_w(\bk)],
\end{equation}
where $\tilde{s}_\rho(\bk)$ and $\tilde{s}_w(\bk)$ are the Fourier transforms of the mass smoothing window $s_\rho(\bx)$ and kernel sampling window $s_w(\bx)$, respectively.  The mass smoothing window takes into account the CIC mass assignment scheme for constructing the coarse density field.  The kernel sampling window corrects for the fact that the long-range kernel given by equation (\ref{eqn:wl}) is sampled on a coarse grid.  In Fourier space, the corrected short-range potential kernel is now given by
\begin{equation}
\label{eqn:wscorrected}
\tilde{w}_s(\bk)=\tilde{w}(\bk)-\tilde{w}_l(\bk)\tilde{s}_\rho(\bk)\tilde{s}_w(\bk),
\end{equation}
and can be slightly anisotropic, particularly near the short-range cutoff.

\begin{figure}[t]
\begin{center}
\includegraphics[width=3in]{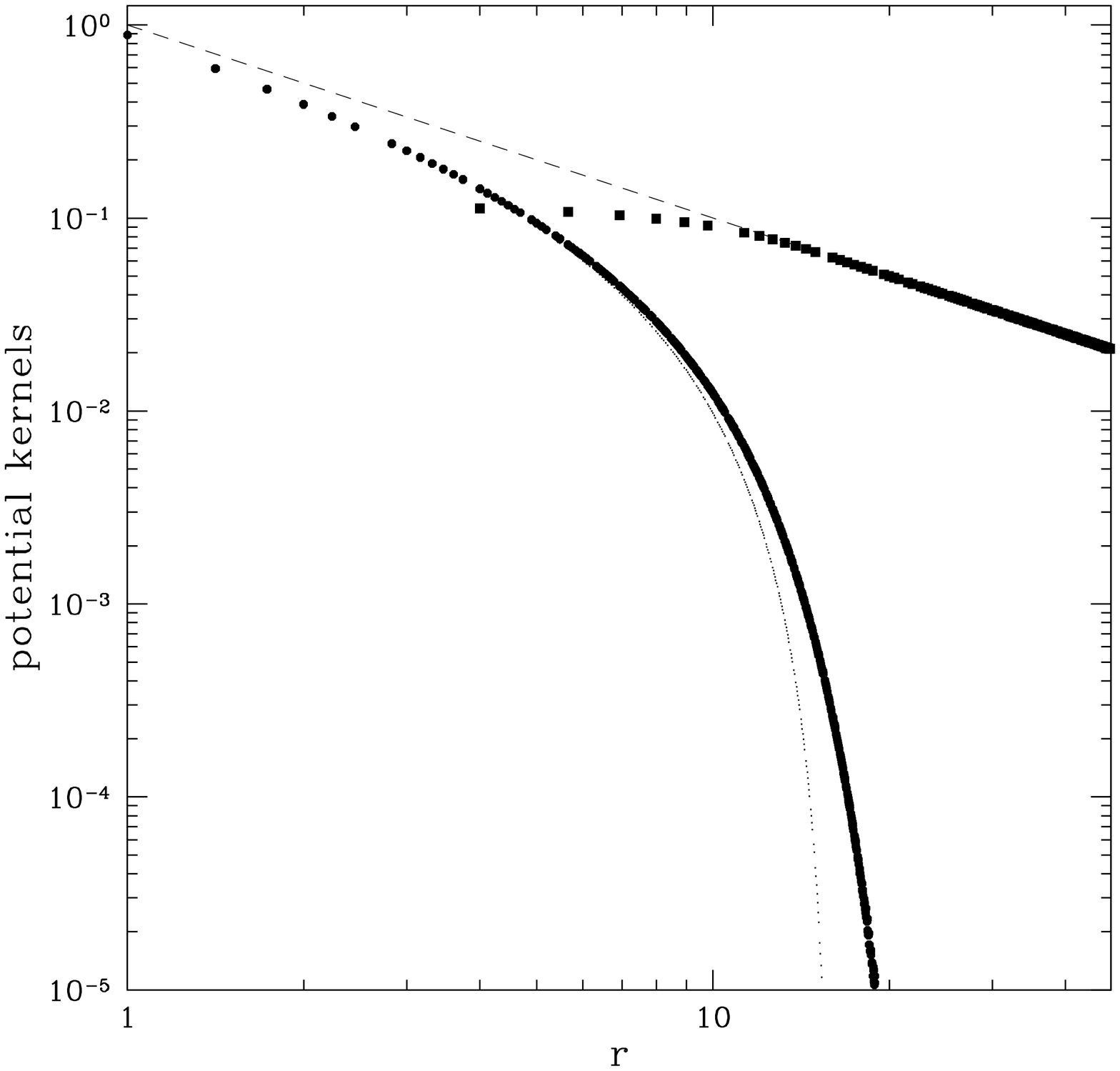}
\includegraphics[width=3in]{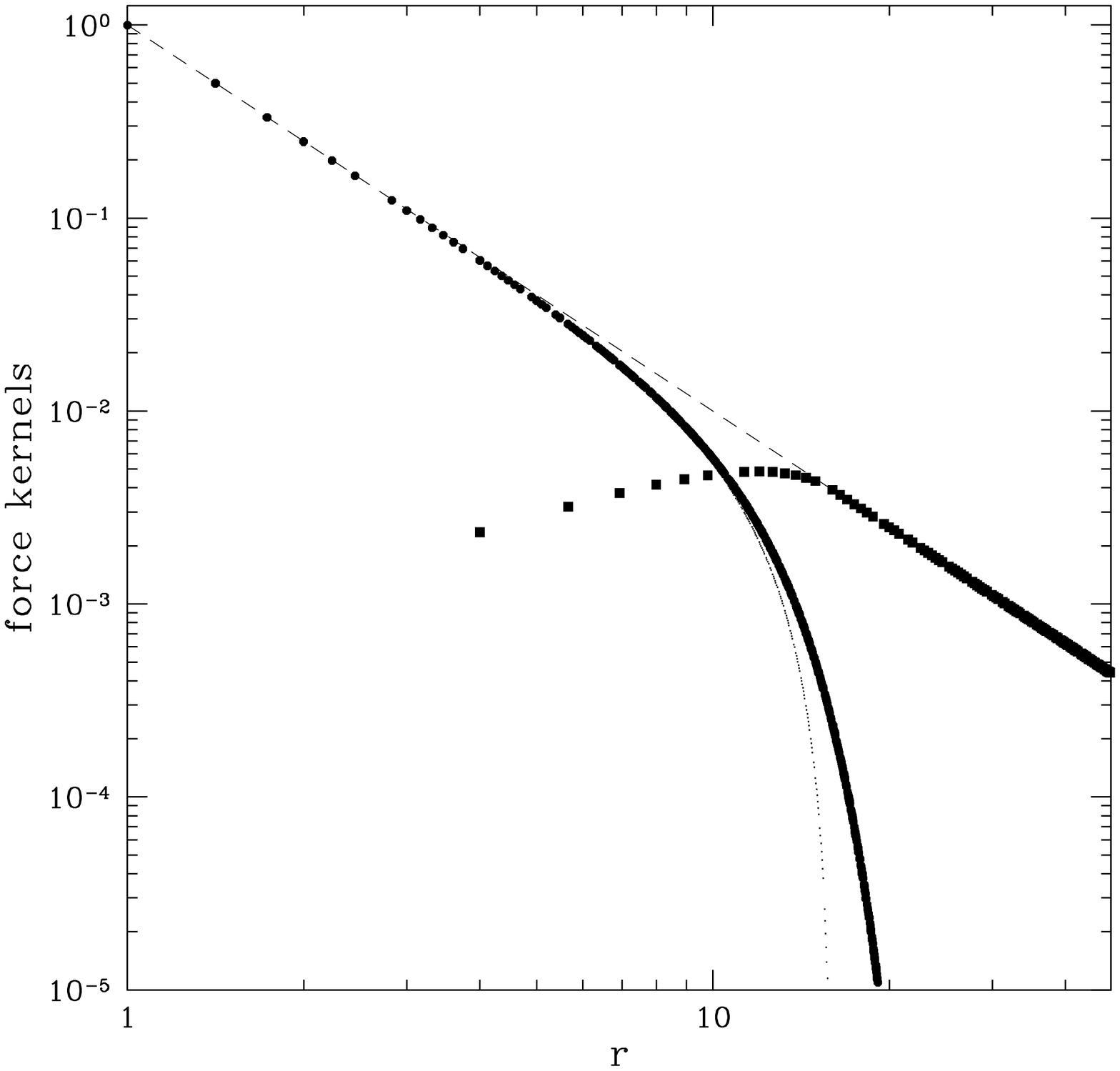}
\begin{alltt}
{\bf \hspace{0.3in}(a)\hspace{3.0in}(b)}
\end{alltt}
\end{center}
\caption[Kernels for the Two-level Mesh Gravity Solver]{The decomposition of the potential kernel (a) and radial force kernel (b) for the two-level mesh gravity solver.  The long-range (filled squares), uncorrected short-range (small dots), and corrected short-range (filled circles) components are plotted.  For comparison, the $1/r$ isotropic potential kernel and the $1/r^2$ isotropic force kernel are illustrated by the dashed lines.  Here, the uncorrected short-range cutoff is $r_g=16$ fine grid cells and the corrected cutoff is approximately 20 grid cells.}
\label{fig:kernels}
\end{figure}

Plotted in Figure \ref{fig:kernels}(a) are the various components of the potential kernel for an initial short-range cutoff $r_c=16$ fine grid cells.  For comparison, the isotropic kernel $w(r)$ is illustrated by the dashed line. The long range kernel $w^c_l(\bx)$ is constructed on a coarse grid and the values in each cell are plotted with filled squares.  The uncorrected and corrected short-range kernels $w^f_s(\bx)$ are constructed on a fine grid and plotted with small dots and filled circles, respectively.  The corrected short-range kernel smoothly goes to zero at a radius now slightly larger than $r_c$.  The new gravity cutoff $b_g=20$ fine grid cells sets the size of the buffer for the local blocks.

For each local block in the domain decomposition, the short-range potential $\phi_s^f(\bx)$ is computed by performing the convolution over the high-resolution density field $\rho^f(\bx)$.  The superscript $f$ denotes that the local fields are constructed on a fine grid.  The short-range force field $\bforce_s^f(\bx)$ is then obtained by finite differencing the short-range potential.  

Note that as an alternative, the force can be calculated directed using the convolution,
\begin{equation}
\bforce_i(\bx)=\int\rho(\bx')w_i(\bx-\bx')d^3x',
\end{equation}
where the anisotropic force kernels are given by
\begin{equation}
w_i(\bx)=-G\frac{x_i}{r^3}.
\end{equation}
The force kernels can be decomposed into short and long range components in a similar fashion to that described previously for the potential kernel.  In Figure \ref{fig:kernels}(b), the magnitude of the force kernels are plotted for a short-range cutoff $r_g=16$ fine grid cells.

In the potential method, the finite differencing degrades the pair-wise force resolution by a few grid cells, but the net force on any given cell or particle is still highly accurate in general.  The force method exactly reproduces the pair-wise inverse-square law on the grid, but not for particles since the CIC mass assignment scheme smoothes the mass and pair-wise force near the grid scale.  Both the potential and force methods require one FFT to forward transform the density field, but the potential method only requires one inverse transform for the potential field while the force method requires three inverse transforms for the force components.  Hence, the force method comes at the cost of two extra FFTs.  For the high-resolution short-range force field, the relatively small accuracy trade-off of the potential method is preferred over the factor of 2 increase in computational work of the force method.  For the lower-resolution long-range force field, the force method allows us to avoid finite differencing the coarse grid and further degrading the resolution of an already smoothed field.  We can afford the extra factor of 2 increase in cost since the long-range force calculation is already a factor of 64 cheaper than the collective short-range force calculations.  

In summary, we combine the potential and force methods to generate a two-level mesh gravity solver that is accurate, cost-effective, and memory-efficient.  The short-range force is computed using the potential method while the long-range force is solved directly using the force method.  We choose an initial short-range cutoff $r_c=16$ fine grid cells and a gravity cutoff $b_g=20$ fine grid cells.  A version of this two-level mesh gravity solver is used by \citet*{2004MerzPMFAST}.

\subsubsection{Pair-wise force test}

\begin{figure}[t]
\center
\includegraphics[width=5in]{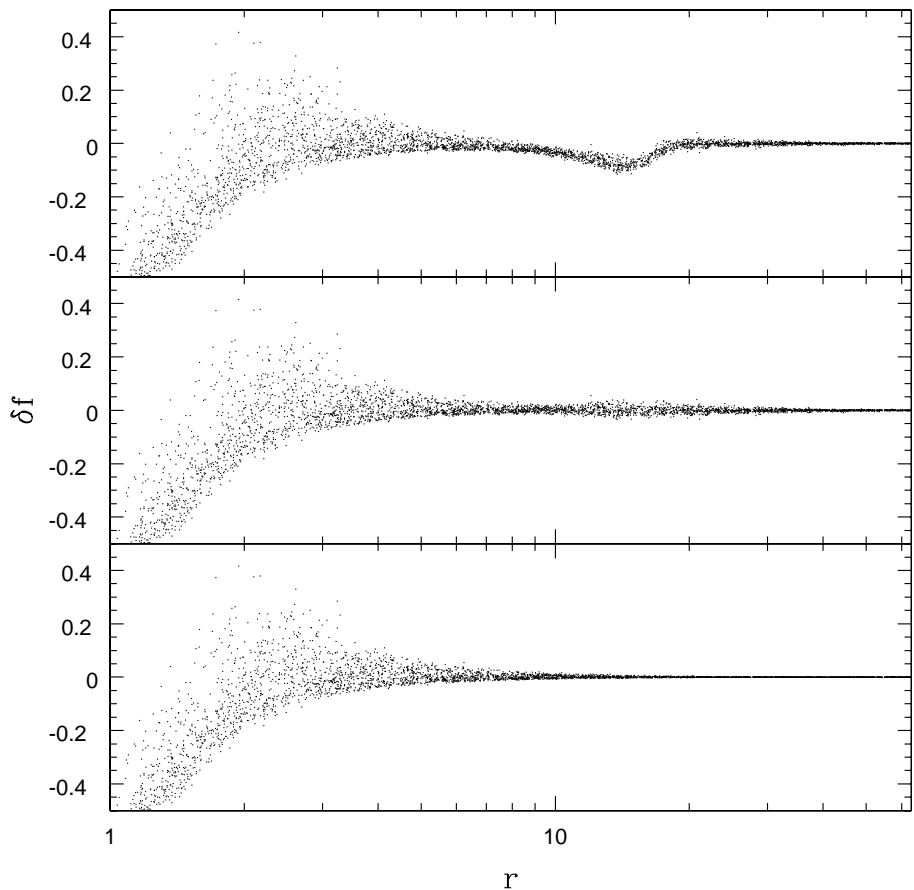}
\caption{Fractional errors for pair-wise force test.  The top and middle panels show the errors for the uncorrected and corrected two-level mesh gravity solver with $r_g=16$ grid cells, while the bottom panel shows the errors for a standard single-mesh solver.  With the corrected kernel, the mean error for separations in the range $r=r_g\pm4$ is zero, the rms error is $<2\%$, and the maximum error is $<5\%$.}
\label{fig:pairwisetest}
\end{figure}

The accuracy of the two-level mesh gravity solver can be checked by performing a pair-wise force test.  We place a particle randomly in a box, calculate both the short and long range force fields, and measure the pair-wise force for a number of randomly placed test particles.  We choose 128 random centers and for each center, $10^4$ test particles are randomly distributed with equal numbers per radial bin to quantify the errors.  In Figure (\ref{fig:pairwisetest}) we plot the fractional error
\begin{equation}
\delta f\equiv\frac{|\bforce|-f(r)}{f(r)},
\end{equation}
where $f(r)=G/r^2$ is the magnitude of the exact force at separation $r$.  The top and middle panels show the fractional errors for the uncorrected and corrected two-level mesh gravity solver while the bottom panel shows the errors for a standard one-level mesh gravity solver based on the potential method.  The same random subset of pairs, with equal numbers per logarithmic bin, are plotted for all three cases for proper comparison.

In all three cases, the fractional errors can approach unity near the grid scale and the source of the problem is the CIC mass assignment scheme.  The cubical cloud shape softens the mass resolution and introduces force anisotropies.  In principle, higher-order mass assignment schemes \citep{1988HockneyEastwood} like triangular-shaped clouds (TSC) can reduce the force fluctuations near the grid scale.  However, \citet{1985Efstathiou} have conducted a detailed study of the PM N-body method and have concluded that the CIC scheme offers the best balance between computational cost and force accuracy.

Near the initial cutoff $r_c=16$ grid cells, the magnitude of the force is under predicted by the uncorrected gravity solver.  The mass smoothing and kernel sampling on the coarse grid are responsible for the decrease in power.  After correcting for these effects, the mean error for separations in the range $r=r_c\pm4$ is zero, the rms error is less than 2\%, and the maximum error is less than 5\%.

\section{Out-of-core Computation}
\label{sec:computation}

At the Canadian Institute for Theoretical Astrophysics (CITA) we have a HP/Compaq GS320 Alpha server with 32 processors and 64 GB of shared memory.  The total floating-point speed has a peak value of 48 Gflops and the total memory bandwidth is 14 GB/s.  The server has a total of 8 independent PCI busses, each running at 266 MB/s.  In the current configuration, an array of 84 SCSI disks providing 3 TB of storage is connected to the server through 6 high bandwidth dual Ultra3 SCSI controllers, each with a peak bandwidth of 320 MB/s.  To improve I/O performance, we stripe the disk array using the HP/Compaq Logical Storage Manager software package.  We can achieve read and write speeds of up to 500 MB/s when direct I/O is used.

Out-of-core computation requires dividing the global data domain into smaller local blocks that can fit entirely in memory.  We work on each block sequentially:  a data block is read from disks into memory, processed in parallel for a number of time steps, and then written back to the disk array.  In the following sections, we describe the domain decomposition, time stepping, and optimizations for doing out-of-core computation.

\subsection{Domain Decomposition}

The three-dimensional global physical domain is decomposed into cubical regions and each local region is extended by buffer regions from its 26 neighbours.  Associated with each extended local region is a high resolution hydro array and a dark matter particle list.  The collisionless dark matter particles only interact gravitationally and for each gravity step, a buffer length of $b_g=20$ grid cells is required to calculate the short-range force without introducing artificial boundary effects.  The moving frame hydro algorithm requires 6 buffer cells per single time step and a hydro buffer $b_h=12$ cells for a double time step.  A total buffer size of $b_t=32$ grid cells is required for every double time step.  Only one gravity step is done per double hydro step.

The periodic global domain is represented solely by a total density field, constructed on a lower-resolution mesh.  The global density field is used to calculate the long-range force field and it can also be used to provide another level of buffering for solving the short-range force field.  To do multiple time steps while the data is in memory, one can increase the size of the buffer, but this is inefficient.  A data block contains both the local physical region plus buffers and the maximum allowable size is limited by the amount of memory in the system.  If the buffer size is increased, then the size of the local region must decrease, resulting in extra work overhead when trying to solve a fixed-size problem.  We can do an additional short-range gravity step by using the global density field to construct an additional buffer for the high-resolution local density field.

\subsection{Time Stepping}

\begin{figure}[t]
\begin{alltt}
{\sf 
call long\_range\_force

do c=1,Nc
   do b=1,Nb
      do a=1,Na
         call read\_data
         call long\_range\_acceleration
         call hydro\_forward
         call pm\_gravity
         call hydro\_reverse
         call hydro\_forward
         call pm\_gravity
         call hydro\_reverse
         call write\_data
      enddo
   enddo
enddo
}
\end{alltt}
\caption{A schematic Fortran code illustrating the time stepping in the OCH cosmological code.  We loop over the decomposed domain and work on each local block sequentially:  a data block is read from disks into memory, processed in parallel for a number of time steps, and then written back to the disk array.}
\label{fig:ochcode}
\end{figure}

A schematic Fortran code is presented in Figure \ref{fig:ochcode} to illustrate the time stepping in the code.  At the beginning of each stepping iteration, we read the global density field from the disks and compute the long-range force field.  The global data is stored on a lower-resolution mesh and fits entirely in memory.  The global long-range force field is divided into local blocks and written back to the disks.  This set of tasks is performed by the subroutine ${\sf long\_range\_force}$.  The global physical domain is decomposed into $N_a\times N_b\times N_c$ number of local regions and we work on each sequentially.  The subroutine ${\sf read\_data}$ creates an extended local block by reading local and buffer data from files on disks.  We then update the velocities of the gas and dark matter particles using the long-range force field.  This is done with the subroutine ${\sf long\_range\_acceleration}$.

A forward hydro sweep, gravity step, and reverse hydro sweep is then performed with the subroutines ${\sf hydro\_forward}$, ${\sf pm\_gravity}$, and ${\sf hydro\_reverse}$, respectively.  In the gravity step, we construct a high resolution total density field from the synchronized gas and dark matter.  The extended local block already contains a buffer of length $b_t$.  In addition, the density field has another level of buffering drawn from the global density field.  The gas velocities and the dark matter particle velocities and positions are advanced by two timesteps in the gravity step.  In our multi time step scheme, the short-range tasks are repeated again.  

The data for the next stepping iteration is prepared in the subroutine ${\sf write\_data}$.  A new local block and new buffer blocks are written to individual files.  In addition, the gas and dark matter are used to construct a portion of the lower resolution global density field to be used by the subroutine ${\sf long\_range\_force}$ in the next stepping iteration.

In total for each stepping iteration, the gas and dark matter are advanced by 4 time steps while the data is in memory.  To be second-order accurate, the value of the time step is fixed during a double step and also between the last double step and the first double step of the next stepping iteration.  The first criterion is imposed by the hydro algorithm and the short-range acceleration while the the latter criterion is required by the long-range acceleration.  The value of the time steps between consecutive double steps in each stepping iteration can be different.

\subsection{Optimizations}

The OCH code is optimized in several ways.  In order to be feasible, out-of-core computation must avoid being I/O limited.  We can significantly reduce I/O overhead by performing disk operations concurrently with numerical calculations.  The first local block of data is read into memory and while computation is being done on it, we simultaneously read in the second local block.  While doing computation on the second local block, we write out the first block and read in the third block.  One thread in our multi-processor machine is assigned to doing I/O.  With this scheme, we have managed to effectively hide the disk operations and reduce I/O overhead to less than 10\%.

We have also significantly reduced the amount of disk operations with the two-level mesh gravity solver and the multi-stepping scheme.  The two-level mesh gravity solver allows FFTs to operate on data stored entirely in memory, which is much faster than the alternative of computing the transforms out-of-core through numerous non-sequential disk accesses.  The multi-stepping scheme allows us to advance the gas and dark matter by 4 time steps before writing the data back to disks.

\section{Cosmological Initial Conditions}

Cosmological initial conditions for large simulations are also generated out-of-core using a two-level mesh scheme \citep{1997PenIC}.  In the standard method, the overdensity field $\delta(\bx)$ is constructed from the convolution, 
\begin{equation}
\delta(\bx)=\int n(\bx')w(\bx-\bx')d^3x',
\end{equation}
of a random white noise field $n(\bx)$ with a density kernel $w(\bx)$, whose Fourier transform is the square root of the initial matter power spectrum $P_i(k)$.  In the two-level mesh scheme, the density kernel is decomposed into short-range (Eq.~[\ref{eqn:ws}]) and long-range (Eq.~[\ref{eqn:wl}]) components.  The density cutoff is denoted $r_d$ to differentiate it from the gravity cutoff $r_g$.  In this case, the function $\alpha(r)$ is chosen to be
\begin{equation}
\alpha(r)=(1+ar^2+br^4+cr^6)w(r),
\end{equation}
with coefficients,
\begin{equation}
\begin{aligned}
&a=-\frac{3}{r_d^2},\\
&b=\frac{3}{r_d^4},\\
&c=-\frac{1}{r_d^6},
\end{aligned}
\end{equation}
that satisfy the conditions given by equation (\ref{eqn:kernelconditions}).  This parametrization is more generic in that the coefficients are independent of the global density kernel $w(r)$ and its derivatives.

After the matter density field has been determined, the gravitational force field is calculated and used to construct the comoving displacement field
\begin{equation}
\bdx(\bx)=-\frac{1}{4\pi G\bar{\rho}}\boldsymbol{\grad}\Phi,
\end{equation}
and the proper peculiar velocity field
\begin{equation}
\bvel_p(\bx)\equiv a\frac{d\bx}{dt}=a\frac{\dot{D}}{D}\bdx,
\end{equation}
where $\bar{\rho}$ is the comoving mean density, $D(t)$ is the linear growth factor, and $\dot{D}$ is its time derivative.  The dark matter particles are displaced from a uniform distribution and their velocities are determined by interpolating from the grid.  The gas distribution is taken to trace the matter distribution and its initial conditions are readily obtained from the grid-constructed fields.  The Poisson solver in the two-level mesh cosmological initial conditions generator is based on the force method, which can calculate the gravitational force field exactly for the grid-constructed matter density field. 

The two-level cosmological initial conditions generator is tested with the following exercise.  Consider a periodic simulation box of comoving length $L=100h^{-1}$ Mpc that is discretized by a $512^3$ grid.  The global domain is divided into $2\times2\times2$ number of cubical local regions.  Each local block  is extended by buffers and has a length of $256+2b_t$ grid cells.  The density cutoff and gravity cutoffs are chosen to be $r_d=30$ and $r_g=28$ grid cells, respectively, and a total buffer size $b_t=32$ grid cells is needed after correcting the short-range kernels using equation(\ref{eqn:wscorrected}).  The gas is discretized on the $512^3$ grid and the dark matter is represented by $256^3$ particles.

\begin{figure}[t]
\begin{center}
\includegraphics[width=3in]{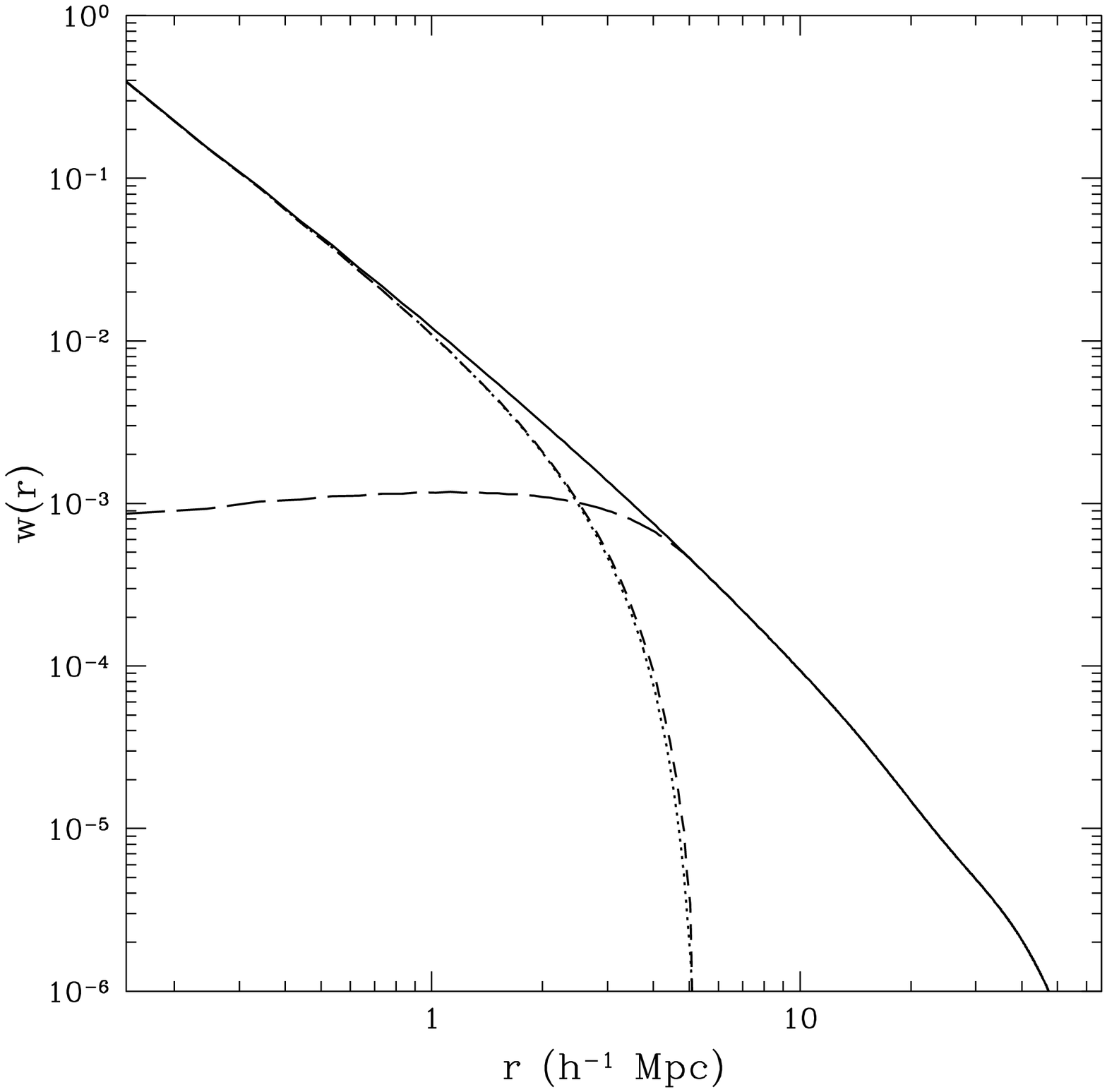}
\includegraphics[width=3in]{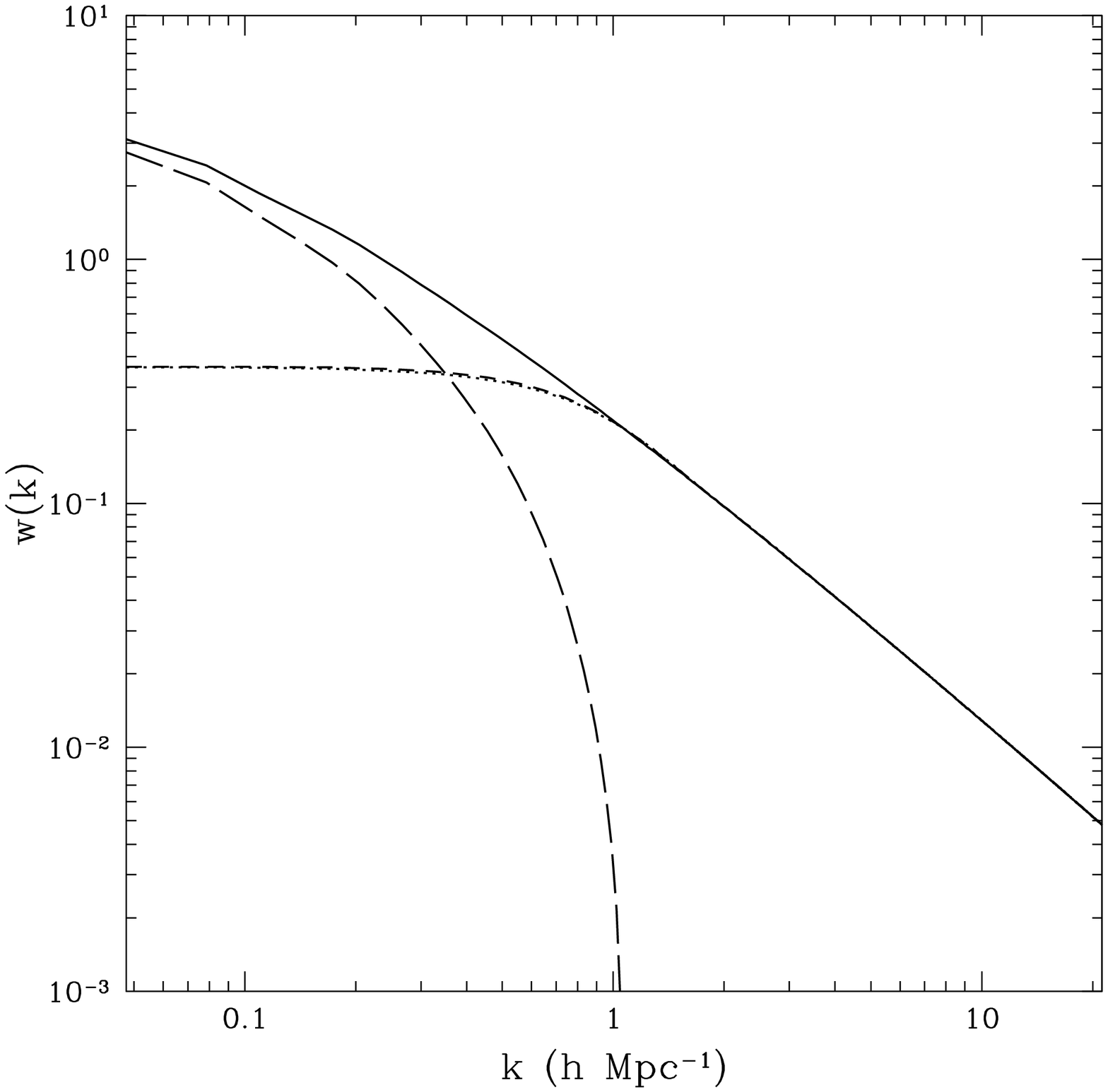}
\begin{alltt}
{\bf \hspace{0.3in}(a)\hspace{3.0in}(b)}
\end{alltt}
\end{center}
\caption{The decomposition of the density function (solid lines) for the two-level mesh initial conditions generator.  The uncorrected short-range (dotted), corrected short-range (short dash), and long-range (long dash) components are plotted.  The real space functions with a short-range cutoff $r_g=30$ fine grid cells are shown in (a) and the Fourier space versions are in (b).}
\label{fig:densityfnc}
\end{figure}

In Fourier space, the isotropic density function $w(k)$ is taken to be the square root of the initial matter power spectrum $P_i(k)$.  The matter transfer function is computed using CMBFAST \citep{1996SeljakZaldarriagaCMBFAST} and with WMAP parameters:  $\Omega_m=0.27$, $\Omega_\Lambda=0.73$, $\Omega_b=0.044$, $n=1$,  $\sigma_8=0.84$, and $h_0=0.7$ \citep{2003SpergelWMAP}.  The initial conditions are generated for an initial redshift $z=50$ where the matter power spectrum is still linear and the real space density field has $|\delta_\text{max}|<1$.  The density function $w(k)$ is integrated to obtain the real space density function $w(r)$, which is then decomposed into short and long range terms for the two-level generator.  The decomposition is shown in Figure \ref{fig:densityfnc}.  For $r_d=30$ grid cells, the corrected and uncorrected short-range terms are very similar.  The correction is only noticeable on scales very near the cutoff.

\begin{figure}[t]
\begin{center}
\includegraphics[width=3in]{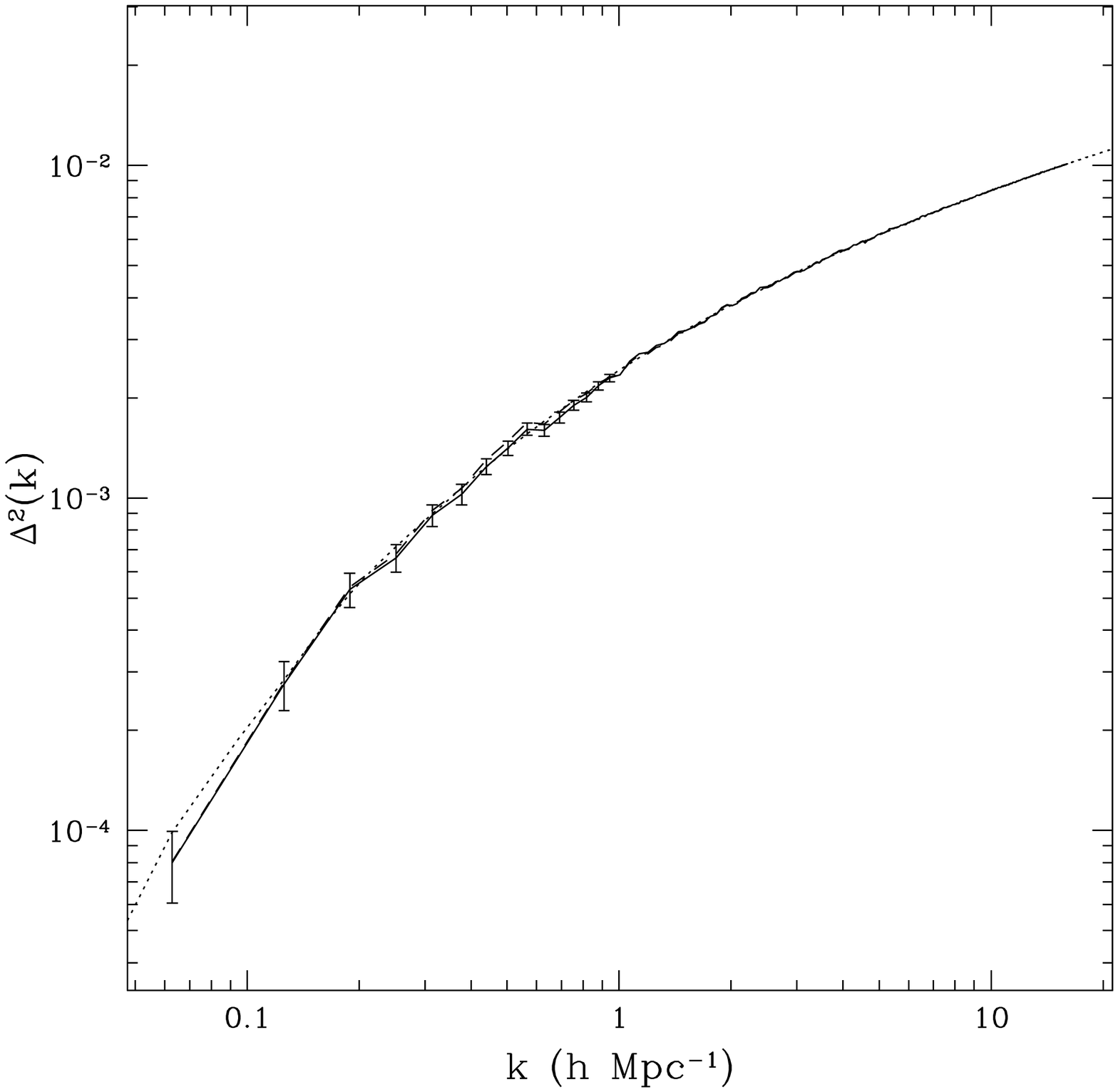}
\includegraphics[width=3in]{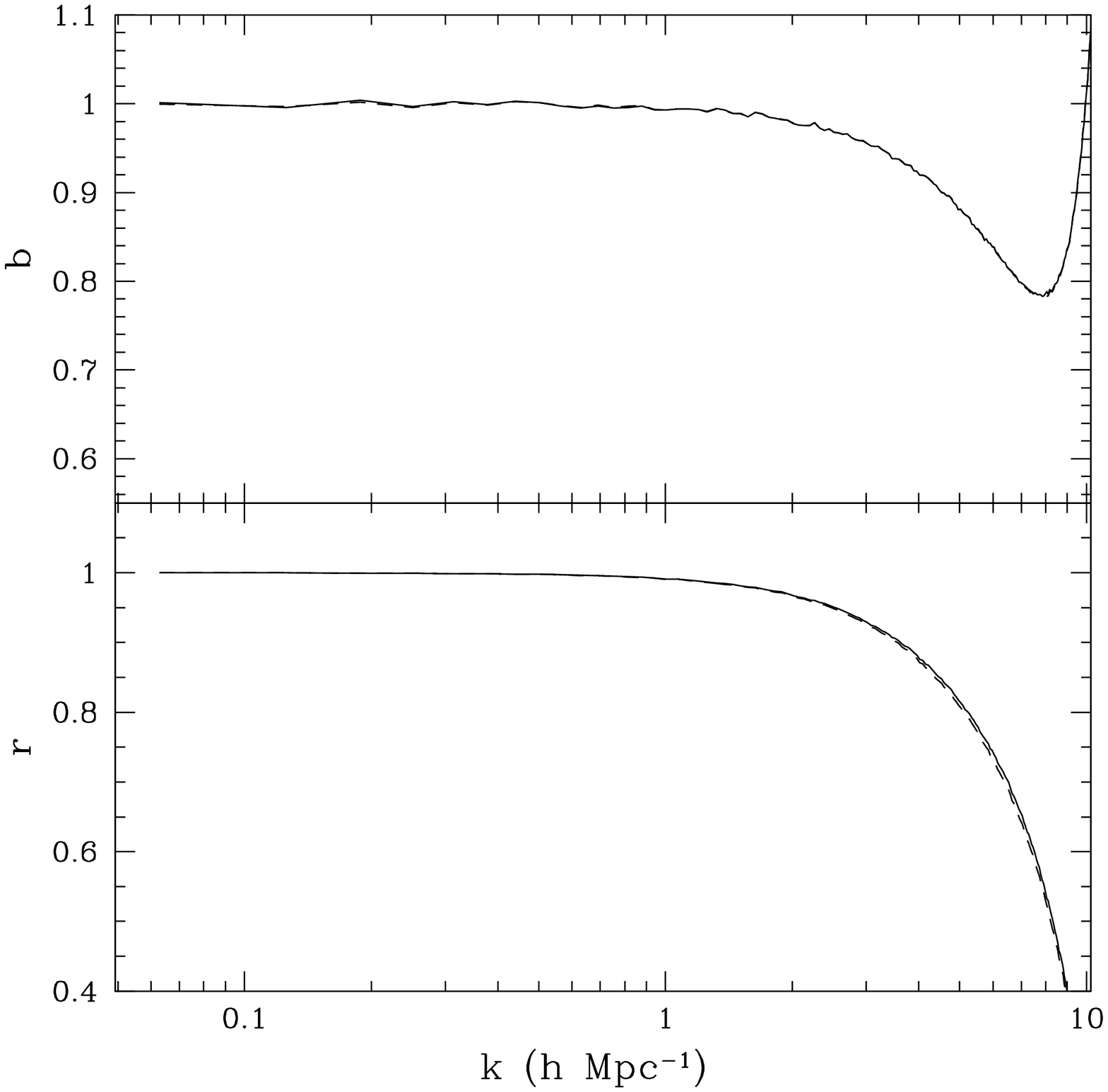}
\begin{alltt}
{\bf \hspace{0.3in}(a)\hspace{3.0in}(b)}
\end{alltt}
\end{center}
\caption{Sample realization from the two-level mesh cosmological initial conditions generator.  In (a), the realization power spectrum (solid line) of the grid-constructed matter density field is plotted against the template CMBFAST power spectrum (dotted line), out to the grid Nyquist frequency.  In (b), the bias and cross-correlation between the dark matter particles and the matter density field are plotted out to the mean interparticle spacing.  The two-level mesh generator agrees with a standard one-level version (dashed lines) to 1\% at all scales.}
\label{fig:pkinit}
\end{figure}

Plotted in Figure \ref{fig:pkinit}(a) is the power spectrum from a sample realization of the initial matter power spectrum.  The dimensionless power spectrum is defined as
\begin{equation}
\Delta^2(k)=\frac{L^3}{2\pi^2}k^3P(k),
\end{equation}
where $k=2\pi/\lambda$.  The realization power $P(k)$ is calculated using FFTs and averaged in bins that span the range $k\pm\Delta k/2$ and have widths $\Delta k=2\pi/L$.  For a Gaussian random field, the fractional error in the realization power spectrum is given by
\begin{equation}
\frac{\Delta P}{P(k)}\sim\sqrt{\frac{2}{N(k)}},
\end{equation}
where $N(k)$ is the number of sampling modes for wavenumber bin $k$.  In a finite-volume periodic box, we have $N(k)\propto k^2$.  The grid-constructed matter density field (solid) has a power spectrum that statistically matches the template CMBFAST spectrum (dotted).  At large wavenumbers or small scales, the agreement is very good due to the high number of sampling modes per bin.  The deviations at small wavenumbers or large scales are expected due to sample variance. 

Figure \ref{fig:pkinit}(b) compares the realization grid power spectrum $P_{gg}(k)$ and the realization particles power spectrum $P_{pp}(k)$, out to the mean interparticle spacing.  To measure the dark matter power spectrum, the $256^3$ particles are mapped onto the $512^3$ grid using the CIC mass assignment scheme.  The statistical correlation between the particles and grid can be quantified with the bias
\begin{equation}
b(k)\equiv\sqrt{\frac{P_{pp}(k)}{P_{gg}(k)}},
\end{equation}
and cross-correlation
\begin{equation}
r(k)\equiv\frac{P_{gp}(k)}{\sqrt{P_{gg}(k)P_{pp}(k)}},
\end{equation}
where $P_{gp}(k)\equiv\langle\delta_g(k)\delta_p(k)\rangle$ is the cross spectrum.   The cross-correlation coefficient or stochasticity parameter can have values $-1\leq r\leq 1$.  On large scales, the particles are perfectly correlated with the grid and unbiased.  On small scales, finite resolution reduces the correlation and power.  The turnover at small scales arises mainly because of the CIC mass assignment scheme.  

The two-level initial cosmological conditions generator (solid lines) can be directly compared to a standard one-level version (dashed lines) by running the latter on the same noise field used in the former.  For both the grid-constructed matter density field and the particle density field, the two-level generator agrees with the standard one-level version to 1\% at all scales.

\section{Cosmological Simulations}

\begin{figure}
\center
\includegraphics[width=5in]{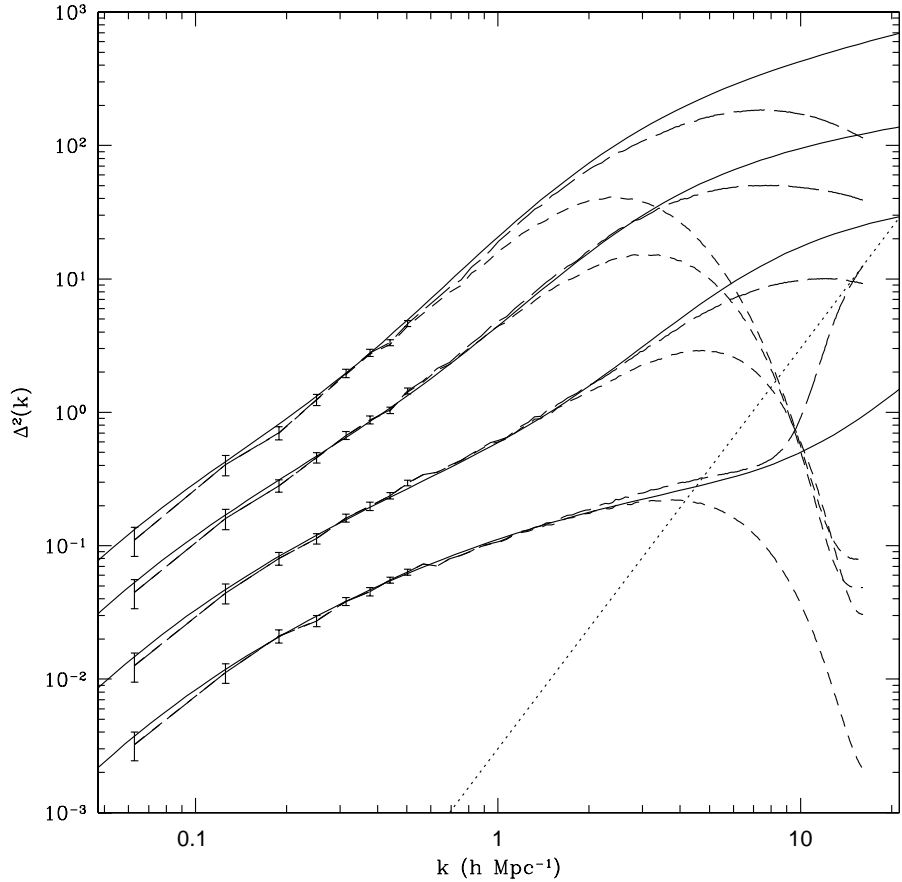}
\caption{Nonlinear evolution of the dark matter (long dashed lines) and gas (short dashed lines) power spectra at redshifts $z=0$, 1, 3, and 7 from an out-of-core simulation with $512^3$ grid cells and $256^3$ dark matter particles in a 100 $h^{-1}$Mpc box.  The simulated dark matter power spectra agree with the matter power spectra (solid lines) predicted using the fitting functions of \cite{2003SmithPeacock}.  Poisson noise have been subtracted from the dark matter power spectra, except at $z=7$ where the power does not exceed the shot noise power (dotted line).}
\label{fig:ochpk}
\end{figure}

The OCH cosmological code is applied to evolving the sample initial conditions generated in the previous section to demonstrate that it accurately simulates nonlinear structure formation in the universe.  The configuration where the global domain is decomposed into $2\times2\times2$ number of cubical blocks used in the generation of the initial conditions is retained for the out-of-core simulation.  The dynamical evolution is checked by measuring the gas and dark matter mass power spectra at redshifts $z=0$, 1, 3, and 7.  Power spectra for the periodic density fields are computed using FFTs.  For each redshift, the 8 local blocks are combined to construct the complete gas density field on a $512^3$ grid.  The clustered distribution of $256^3$ dark matter particles is mapped onto a higher resolution $1024^3$ grid to improve the accuracy of the measurements.

In Figure \ref{fig:ochpk} the simulated power spectra are plotted against the nonlinear fitting functions of \citet{2003SmithPeacock}.  Note that we use CMBFAST to compute the initial transfer function while \citet{2003SmithPeacock} use that of \citet*{1992EBW} and therefore, some differences are expected.  The simulated dark matter and predicted matter spectra are in good agreement, except at scales near the box size and the grid spacing.  At redshift $z=0$, the comparison is good on frequencies $k\lesssim3h$ Mpc$^{-1}$ or comoving wavelengths $\lambda\gtrsim2h^{-1}$ Mpc.  More than half the power is lost on scales less than 5 grid cells because the force resolution in the PM scheme is softened near the grid scale.  At large scales approaching the box size, the evolution is consistent with linear growth and the differences are due to sample variance in the initial conditions.  The observed deviations are expected since the realization initial dark matter power spectrum has a deficit in power at large scales to begin with (see Figure \ref{fig:pkinit}).  On linear and moderately nonlinear scales, the simulated gas and dark matter are highly correlated with no bias.  On nonlinear scales, the gas loses power because the pressure prevents gravitational collapse and the force softening of the dark matter.  The two-level mesh out-of-core results are very similar to that produced with the Hydro\&N-body code of \citet{2004TracPenMACH}, differing by less than 5\% across the entire wavelength range and some deviation is expected because of the Poisson noise.

\subsection{Code comparison}

\begin{figure}[t]
\begin{center}
\includegraphics[width=3in]{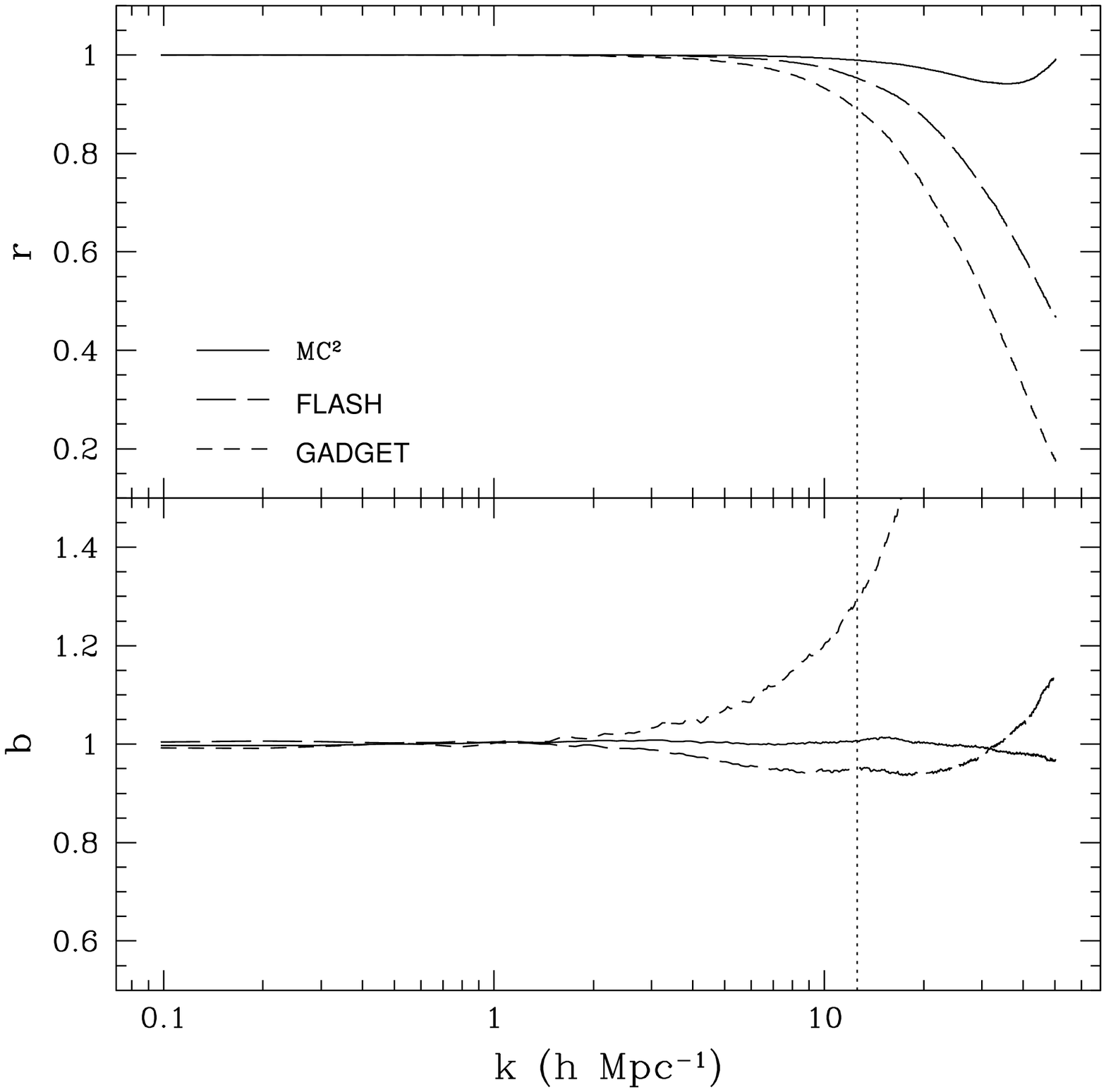}
\includegraphics[width=3in]{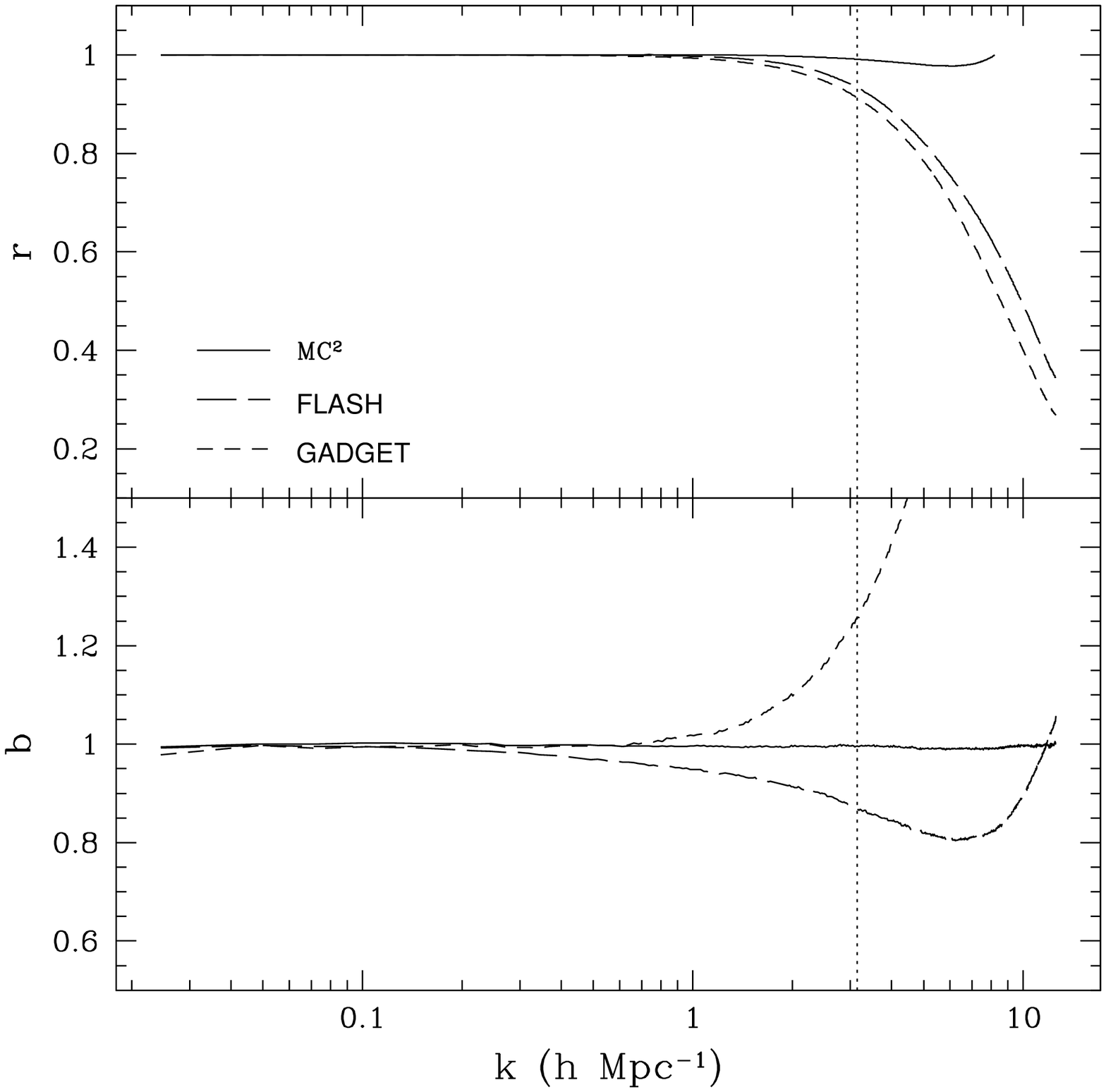}
\begin{alltt}
{\bf \hspace{0.3in}(a)\hspace{3.0in}(b)}
\end{alltt}
\end{center}
\caption{The bias and cross-correlation with respect to the PMM results for box sizes $L$ of 90 Mpc (a) and 360 Mpc (b), respectively.  The dotted line represents the particle Nyquist frequency.  The grid codes PMM, MC$^2$, and FLASH had comoving grid spacings of $L/1024$ while the tree code GADGET used a comoving softening length of 20 kpc.}
\label{fig:codecomparison}
\end{figure}

\citet{2004Heitmann} compared several N-body codes and have found that, in general, the codes show agreement over a wide range of scales at the 5\% level or better.  The comparison included the FFT-based grid code MC$^2$, the AMR code FLASH \citep{2000FryxellFLASH}, the adaptive P$^3$M code Hydra \citep{1995CouchmanHydra}, and the tree codes HOT \citep{1993WarrenSalmon}, GADGET \citep{2001SpringelGADGET}, and TPM \citep{2003BodeOstrikerTPM}.  We have taken the publicly-available initial conditions for two cosmological simulations of a $\Lambda$CDM universe
with box sizes of 90 Mpc and 360 Mpc ($h=0.71$), respectively, and ran them with the N-body algorithm used in the OCH code.  We will refer to this N-body code as the particle-multi-mesh (PMM) code and an application of the code is presented in \citet{2005McDonald}.

The simulations were run from redshift $z=50$ down to $z=0$ with $256^3$ particles on an effective grid of $1024^3$ cells.  The grid cell spacing was chosen to match that used by the grid codes MC$^2$ and FLASH.  In Figure \ref{fig:codecomparison}, our $z=0$ results are compared with those of MC$^2$ and FLASH, as well as with the higher resolution results of GADGET.  The particles are mapped onto a 1024$^3$ grid using the NGP assignment scheme, Fourier transformed using FFTs, and power spectra are calculated while taking into account shot noise.  The bias $b(k)$ and cross-correlation $r(k)$ are measured with respect to the PMM results.  While the PMM, MC$^2$, and FLASH simulations share the same minimum mesh spacing, the former two achieve slightly higher force resolution than the latter.  The differences are probably due to the fact that FLASH only refines in relevant regions and the effective force softening on an adaptive mesh can be higher than that of an FFT-based Poisson solver.  Both the bias and cross-correlation between the PMM and MC$^2$ results are unity up to the particle Nyquist frequency and only deviate by less than 5\% out to the grid Nyquist frequency.  Note that the NGP mass assignment scheme and Poisson noise subtraction are partly responsible for magnifying the differences near the grid Nyquist frequency.  The two FFT-based codes differ in their time-integration schemes, but we suspect they share similar gravity kernels, and the latter is a stronger influence on the dynamical evolution and therefore, the origin of the high agreement in the results.

The PMM simulations were run at lower resolution than the GADGET simulations and the effect of force softening can be seen in Figure \ref{fig:codecomparison}.  At the particle Nyquist frequency, the PMM results show a reduction of $\sim20\%$ in power and the correlation is weakened by $\sim10\%$.  By decreasing the ratio of the grid cell spacing to the mean interparticle spacing, one can further reduce the effects of force softening in a FFT-based gravity solver.  However, this comes at a cost in both work and memory.  In practice, particle-mesh type codes like PMM are more optimal for simulations which require high mass resolution with moderate spatial resolution.

\section{Future Work}

The Beowulf clusters are becoming the dominant HPC systems in computational astrophysics because of the unrivaled speeds that can easily reach into the teraflops.  These systems remain memory-limited, but simulations can be run out-of-core.  IDE hard drives are relatively cheap and tens to hundreds of terabytes of disk space can be added.  Hardware or software raiding can be used to prevent the loss of data in the expected event of disk failures.  For the distributed-memory cluster, the implementation will require a multi-level mesh scheme with parallelization across nodes using the Message Passing Interface (MPI) libraries.  With an out-of-core, distributed-memory code, we will be capable of running hydrodynamic simulations with a trillion grid cells and particles.

\section{Conclusions}

We have developed an out-of-core hydrodynamic code and an out-of-core initial conditions generator for high resolution cosmological simulations that require terabytes of memory.  The OCH code utilizes a two-level mesh gravity solver and a multi-stepping scheme that significantly reduce the amount of disk operations.  In addition, we have managed to reduce I/O overhead down to less than 10\% by performing disk operations concurrently with numerical calculations.  The code is cost-effective and memory-efficient.  It has been demonstrated to accurately simulate the nonlinear structure formation in the universe and will provide high mass resolution for cosmological applications.  The first application of the OCH code involves simulating the high redshift intergalactic medium in a WMAP cosmology (Trac \& Pen 2004 in prep).  We are currently running an out-of-core simulation with $2016^3$ grid cells and $1008^3$ dark matter particles in a $50h^{-1}$ Mpc box.  This high resolution simulation has a comoving grid spacing of $\Delta x=25h^{-1}$ kpc and a dark matter particle mass resolution of $\Delta m=7.7\times10^6h^{-1}M_\odot$.

\section{Acknowledgments}

We thank Chris Loken for the much appreciated help in configuring the hardware and software for the out-of-core shared-memory machine.  We also thank Hugh Merz for discussions on the two-level mesh scheme.  And we appreciate that Katrin Heitmann et al. have made their initial conditions and data publicly available.

% The phrase \citep{Bai92} produces (Bailyn 1992).
% In the phrase \citet{Bai95} Bailyn et al. (1995) appear as a noun.
% Affixes (e.g. Barnes et al. 1976) are produced by the phrase
% \citeaffixed{Barnes et al. 1976}{e.g.}.
% Other options of the harvard package, e.g. \citeyear, are not
% reproduced in New Astronomy.

\bibliographystyle{apj}
\bibliography{astro}

\begin{thebibliography}{24}
\expandafter\ifx\csname natexlab\endcsname\relax\def\natexlab#1{#1}\fi

\bibitem[{{Bagla}(2002)}]{2002Bagla}
{Bagla}, J.~S. 2002, Journal of Astrophysics and Astronomy, 23, 185

\bibitem[{{Bode} \& {Ostriker}(2003)}]{2003BodeOstrikerTPM}
{Bode}, P., \& {Ostriker}, J.~P. 2003, \apjs, 145, 1

\bibitem[{{Couchman}(1991)}]{1991Couchman}
{Couchman}, H.~M.~P. 1991, \apjl, 368, L23

\bibitem[{{Couchman} {et~al.}(1995){Couchman}, {Thomas}, \&
  {Pearce}}]{1995CouchmanHydra}
{Couchman}, H.~M.~P., {Thomas}, P.~A., \& {Pearce}, F.~R. 1995, \apj, 452, 797

\bibitem[{{Dubinski} {et~al.}(2004){Dubinski}, {Kim}, {Park}, \&
  {Humble}}]{2004DubinskiGOTPM}
{Dubinski}, J., {Kim}, J., {Park}, C., \& {Humble}, R. 2004, New Astronomy, 9,
  111

\bibitem[{{Efstathiou} {et~al.}(1992){Efstathiou}, {Bond}, \&
  {White}}]{1992EBW}
{Efstathiou}, G., {Bond}, J.~R., \& {White}, S.~D.~M. 1992, \mnras, 258, 1P

\bibitem[{{Efstathiou} {et~al.}(1985){Efstathiou}, {Davis}, {White}, \&
  {Frenk}}]{1985Efstathiou}
{Efstathiou}, G., {Davis}, M., {White}, S.~D.~M., \& {Frenk}, C.~S. 1985,
  \apjs, 57, 241

\bibitem[{{Fryxell} {et~al.}(2000){Fryxell}, {Olson}, {Ricker}, {Timmes},
  {Zingale}, {Lamb}, {MacNeice}, {Rosner}, {Truran}, \&
  {Tufo}}]{2000FryxellFLASH}
{Fryxell}, B., {Olson}, K., {Ricker}, P., {Timmes}, F.~X., {Zingale}, M.,
  {Lamb}, D.~Q., {MacNeice}, P., {Rosner}, R., {Truran}, J.~W., \& {Tufo}, H.
  2000, \apjs, 131, 273

\bibitem[{{Harten}(1983)}]{1983Harten}
{Harten}, A. 1983, J. Comp. Phys., 49, 357

\bibitem[{{Heitmann} {et~al.}(2004){Heitmann}, {Ricker}, {Warren}, \&
  {Habib}}]{2004Heitmann}
{Heitmann}, K., {Ricker}, P.~M., {Warren}, M.~S., \& {Habib}, S. 2004, ArXiv
  Astrophysics e-prints

\bibitem[{{Hockney} \& {Eastwood}(1988)}]{1988HockneyEastwood}
{Hockney}, R.~W., \& {Eastwood}, J.~W. 1988, {Computer simulation using
  particles} (Bristol: Hilger, 1988)

\bibitem[{{McDonald} {et~al.}(2005){McDonald}, {Trac}, \&
  {Contaldi}}]{2005McDonald}
{McDonald}, P., {Trac}, H., \& {Contaldi}, C. 2005, ArXiv Astrophysics e-prints

\bibitem[{{Merz} {et~al.}(2004){Merz}, {Pen}, \& {Trac}}]{2004MerzPMFAST}
{Merz}, H., {Pen}, U., \& {Trac}, H. 2004, submitted to New Astronomy,
  astro-ph/0402443

\bibitem[{{Pen}(1997)}]{1997PenIC}
{Pen}, U. 1997, \apjl, 490, L127

\bibitem[{{Salmon} \& {Warren}(1997)}]{1997SalmonWarren}
{Salmon}, J., \& {Warren}, M. 1997, in Proceedings of the Eighth {SIAM}
  Conference on Parallel Processing for Scientific Computing

\bibitem[{{Seljak} \& {Zaldarriaga}(1996)}]{1996SeljakZaldarriagaCMBFAST}
{Seljak}, U., \& {Zaldarriaga}, M. 1996, \apj, 469, 437

\bibitem[{{Smith} {et~al.}(2003){Smith}, {Peacock}, {Jenkins}, {White},
  {Frenk}, {Pearce}, {Thomas}, {Efstathiou}, \& {Couchman}}]{2003SmithPeacock}
{Smith}, R.~E., {Peacock}, J.~A., {Jenkins}, A., {White}, S.~D.~M., {Frenk},
  C.~S., {Pearce}, F.~R., {Thomas}, P.~A., {Efstathiou}, G., \& {Couchman},
  H.~M.~P. 2003, \mnras, 341, 1311

\bibitem[{{Spergel} {et~al.}(2003){Spergel}, {Verde}, {Peiris}, {Komatsu},
  {Nolta}, {Bennett}, {Halpern}, \& {Hinshaw}}]{2003SpergelWMAP}
{Spergel}, D.~N., {Verde}, L., {Peiris}, H.~V., {Komatsu}, E., {Nolta}, M.~R.,
  {Bennett}, C.~L., {Halpern}, M., \& {Hinshaw}, G. 2003, \apjs, 148, 175

\bibitem[{{Springel} {et~al.}(2001){Springel}, {Yoshida}, \&
  {White}}]{2001SpringelGADGET}
{Springel}, V., {Yoshida}, N., \& {White}, S.~D.~M. 2001, New Astronomy, 6, 79

\bibitem[{{Strang}(1968)}]{1968Strang}
{Strang}, G. 1968, SIAM J. Num. Anal., 5, 506

\bibitem[{{Trac} \& {Pen}(2003)}]{2003TracPenCFD}
{Trac}, H., \& {Pen}, U. 2003, \pasp, 115, 303

\bibitem[{{Trac} \& {Pen}(2004)}]{2004TracPenMACH}
---. 2004, \newast, 9, 443

\bibitem[{Warren \& Salmon(1993)}]{1993WarrenSalmon}
Warren, M.~S., \& Salmon, J.~K. 1993, in Supercomputing, 12--21

\bibitem[{{Xu}(1995)}]{1995Xu}
{Xu}, G. 1995, \apjs, 98, 355

\end{thebibliography}
%\begin{thebibliography}{14}
%\end{thebibliography}

%\begin{thebibliography}{}
% \harvarditem{Name}{Year}{label}
%\harvarditem{Author}{year}{label}Reference
%\end{thebibliography}

\end{document}